# Three-Dimensional Imaging of High-Density Dislocation Networks and Their Interactions using Multislice Electron Ptychography

*Eegene (Clara) Chung[1], Anand Ithepalli[2], Naomi Pieczulewski[2], Chia-Hao Lee[3], Steven Zeltmann[3], Keun-Yeol Park[4], Celesta S. Chang[4], Debdeep Jena[2,5], David A. Muller[3,6]**

[1]Department of Physics, Cornell University, Ithaca, NY 14853, USA

[2]Department of Materials Science and Engineering, Cornell University, Ithaca, NY 14853, USA

[3]School of Applied and Engineering Physics, Cornell University, Ithaca, NY 14853, USA

[4]Department of Physics & Astronomy, Seoul National University, Seoul, South Korea

[5]School of Electrical and Computer Engineering, Cornell University, Ithaca, NY 14853, USA

[6]Kavli Institute at Cornell for Nanoscale Science, Cornell University, Ithaca, NY 14853, USA

ABSTRACT

Development of novel devices often involve combining functional materials and substrates, each chosen for particular physical properties such as low-loss or thermal conductivity. This, however, can introduce a large lattice mismatch and high density of structural defects that may limit device performance and reliability by trapping charge or creating leakage pathways. Such is the case for integrating superconducting TiN with a low-loss sapphire substrate. Here we resolve and image the resulting misfit, screw, and threading dislocations in epitaxial TiN on sapphire using the depth-sectioning capabilities of multislice electron ptychography. We find misfit dislocations spaced ~1.6 nm apart and track out-of-plane crossings of misfit dislocations as well as their transitions to threading dislocations, details which are obscured in conventional defect-imaging methods. The ability to correlate specific dislocation types and their interactions in extremely small volumes and at high densities is valuable for understanding structure-property relations even in modern, scaled devices.

MAIN TEXT

Microscopic defects and their interactions with charge carriers[1–4] can greatly impact the performance of a broad range of devices in semiconductor, optoelectric, and quantum technologies. In semiconductors, screw and threading dislocations can act as current leakage pathways that may cause shorting or degradation in power devices[5–7]. Recently, identification of dislocation networks formed due to strain relaxation from etched sidewalls and their elimination resulted in the realization of first electrically-pumped continuous-wave deep-ultraviolet semiconductor lasers[8]. In polar materials, dislocations can be sources of trapped charges impeding electrical properties[9–11] or otherwise create depolarizing fields that form dead layers in ferroelectric thin film devices[12]. Dislocation networks can also be engineered to enhance material properties[13–18], demonstrating the diverse effects that dislocations can have on devices.

Since the early direct observations of individual dislocations[19], considerable effort has been put into imaging crystalline defects with higher spatial resolution and extending their characterization into three dimensions (3D). Several approaches have been developed for 3D dislocation imaging, including X-ray-based techniques, electron tomography, through-focal scanning transmission electron microscopy (STEM)[20], and recently, electron ptychography[21,22]. X-ray techniques such as topo-tomography[23], laminography[24], coherent X-ray diffraction[25], and dark-field X-ray microscopy[26] can reconstruct 3D dislocation structures at sub-micron scales, best suited for bulk materials containing relatively low dislocation densities and well-separated defects. In highly-mismatched heterostructures[27], dislocation cores may be separated by only 1-2 nm and the resolution of the forementioned X-ray techniques are insufficient to resolve individual dislocations within such dense networks.

Electron microscopy methods provide substantially higher spatial resolution, but conventional TEM and STEM imaging fundamentally produce two-dimensional (2D) projections of 3D structures. Defects such as edge-type misfit dislocations with in-plane Burgers vectors can be identified using a Fourier analysis on high-resolution TEM (HRTEM)[28,29] or atomic-resolution STEM images[30], or directly imaged by dark-field TEM (DF-TEM)[31]. However, projection through the sample obscures defects with significant out-of-plane character including screw and threading dislocations, as well as larger structures such as columnar grain growths. Moreover, orientation-dependent contrast in techniques like DF-TEM can vanish under certain screw dislocation geometries due to symmetry[31,32]. 3D electron microscopy techniques such as electron tomography[33,34] and through-focal STEM[20] partially overcome projection limitations but introduce other challenges: electron tomography, like its 3D X-ray counterpart, typically requires acquisition of on the order of one hundred projection images over a tilt series to recover depth information, making experiments dose and labor-intensive and limiting throughput. Annular dark-field (ADF) STEM data taken at various focal depths can provide moderate depth-sensitivity over a short (4-8 nm) depth-of-focus[20], but suffers from elongation artifacts[35] and relies on an imaging model that breaks down in samples thicker than ~5 nm due to multiple scattering[36–38].

In contrast, multislice electron ptychography (MEP)[39,40] recovers depth information using the parallax effect from beam divergence along the z-axis while requiring only a single x-y scan of the probe. This avoids a lengthy, and often damaging tilt or through-focal series. In a single scan lasting from 2-15 seconds, the full scattering intensity distribution is recorded for every x-y probe position across the sample[41–44] (Fig. 1a). When a convergent beam is scanned over the sample, atoms closer to the beam focus move across the shadow image formed in the diffraction plane faster than those that are further away, as described in Ref.[38]. This parallax effect is encoded in the diffraction pattern[38] and extracting its depth information has allowed atomic-scale roughness and strain profiles of gate-all-around (GAA) devices in 3D[37]. As shown in Fig. 1a, MEP exploits all experimentally available information – including the central transmitted beam (green disk), scattered Bragg beams (blue disks), and phase modulations within their overlap regions (red) – to iteratively reconstruct the phase of the specimen transmission function, which is proportional to the sample's atomic potential[45]. In the multislice formulation, the three-dimensional electrostatic potential of the sample is divided into a series of thin projected-potential "slices" (Fig. 1a, inset), allowing multiple scattering to be explicitly incorporated into the reconstruction rather than treated as an imaging artifact[39,40]. This approach significantly improves the maximum attainable lateral resolution compared to single-slice ptychography[46] and extends the reconstructable sample thickness well beyond the ~2-5 nm depth-of-field[47] of electron ptychography. As a result, samples up to 40-50 nm thick can now be routinely reconstructed with a lateral resolution up to 0.2 Å and depth resolution of 2-4 nm[37,38,40,48]. Combined with established 2D image analysis techniques such as geometric phase analysis (GPA)[28,29], MEP has enabled 3D characterization of single dislocations as well as strain-mapping in single crystals[21] and nanoparticles[22].

Here, we have extended these capabilities by systematically identifying and mapping multiple dislocation types within an epitaxial heterostructure and revealing their atomic-scale interactions in a high-density dislocation network. We applied MEP to a high quality-factor superconducting nitride resonator[49] used for the development of all-nitride crystalline Josephson junctions in superconducting qubits. A 50 nm thin-film of titanium nitride (TiN) was grown via molecular-beam epitaxy (MBE) on a commercial c-plane sapphire (a-$Al_2O_3$) substrate with growth conditions described in Ref.[49]. Fig. 1b shows the low-magnification ADF-STEM image of the heterostructure and Fig. 1c is a schematic of the different types of dislocations that may exist within the epitaxial TiN film near the substrate. Figs. 1d and e show the high-magnification ADF image and the MEP reconstruction at the interface of TiN and sapphire, viewed in $[11\bar{2}]$ zone of TiN parallel to $[110]$ zone of sapphire. In the ADF image of Fig. 1d, bright atoms corresponding to Ti with a moderately high atomic number (Z = 22) dominates the weak signals from lower-Z elements like N and O due to $\sim Z^{1.6}$ high-angle ADF (HAADF) contrast[50,51]. While N cannot be resolved in the presence of the heavier Ti, O is marginally visible alongside Al. In the MEP reconstruction of Fig. 1e, the brightest atoms are Ti which forms a rocksalt cubic structure with dimmer, but visible, N atoms, reflecting the $\sim Z^{0.6}$ atomic number contrast of MEP[40,52]. Unlike ADF-STEM, this allows MEP to simultaneously resolve light elements like N and O near heavier elements like Al and Ti. The reconstructed potential demonstrates a high signal-to-noise ratio (SNR) image with lateral resolution (information transfer limit) of 0.33 Å and depth resolution of 3.7 nm [Supporting Information Figure. 2]. Fig. 1f is a 3D visualization of the reconstructed atomic potential and the different types of dislocations identified in these images, which will be discussed below. The multislice reconstruction consists of 50 slices with 0.85 nm slice thickness, spanning a ~40 nm-thick sample with ~2.5 nm of vacuum padding on the bottom to reduce wrap-around artifacts and 15 nm × 9 nm lateral field-of-view. Data along with experimental and reconstruction parameters are available in [Supporting Information Section 1 and associated files].

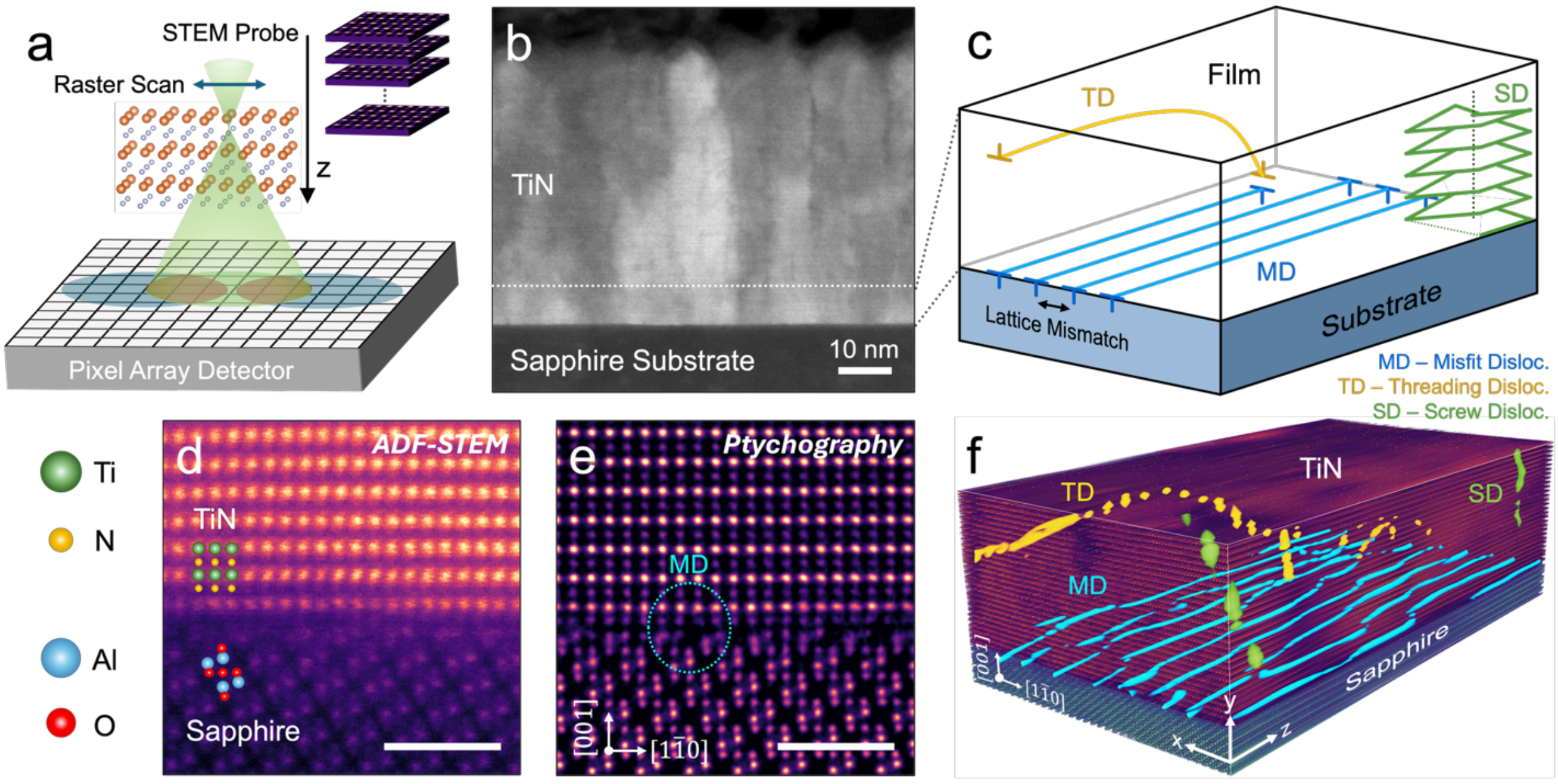

**Figure 1.** a) Schematic of electron ptychography data acquisition using a pixelated detector. b) ADF-STEM image of epitaxial TiN grown on sapphire substrate exhibiting columnar grain growths. Contrast of the columns indicate varying grain orientations relative to the substrate. c) Schematic of the types of dislocations observed in TiN on sapphire. Most commonly present are the misfit dislocations (MD) at the interface of TiN and a-$Al_2O_3$ (sapphire); pure screw dislocations (SD) and mixed-type threading dislocations (TD) are also observed. d) ADF-STEM image of the interface of TiN on sapphire. e) A 0.85 nm-thick slice from the MEP reconstruction at the TiN/sapphire interface. MDs are present at the interface due to the lattice mismatch between TiN and sapphire. f) 3D MEP-measured map of dislocations present in TiN on sapphire. At the interface, a periodic array of MD lines appear as the long cyan-colored strips. Each strip follows the trajectory of a MD through the depth of the sample. Yellow markers map the trajectory of the TD and green markers indicate the lattice distortions from SDs. All scale bars are 1 nm unless noted otherwise.

In Fig. 2, the 3D structure of a screw dislocation (SD) is observed. It shows a high-resolution helical[20] atomic ordering characteristic of the SD. A slice of the MEP reconstruction in Fig. 2a indicates the TiN lattice planes on the righthand side of SD are displaced upwards with respect to the lefthand side. The multislice method disentangles the complex atomic projections of a thick sample into a sequence of thin slices. By accessing each slice, the depth-resolved structure cross-section of a SD viewed side-on is obtained. A 3D visualization about the SD core is shown in Figs. 2b-d. The spiral arrangement of atoms is apparent without any manual postprocessing except for the MEP reconstruction itself; only the brightness and contrast were adjusted to highlight the Ti atoms to simplify the visualization. Due to the coarse z-sampling ($\Delta z = 0.85\ \mathrm{nm/slice}$) relative to that of the lateral sampling ($\Delta x, \Delta y = 0.02\ \mathrm{nm/pixel}$), the depth information was interpolated to preserve a uniform voxel size, hence the lack of atomic resolution along z. Finer z-sampling does not improve the quality of the reconstruction as our depth resolution is limited[40] to 3.7 nm by the scattering geometry. By viewing the SD from the side where we have higher resolution, we have identified the SD core that traverses through the sample at an angle 76.5° with respect to the c-plane of sapphire.

Beyond direct observation, a geometric phase analysis[28–30] is performed to further elucidate the behaviors of the dislocations over the full field-of-view. Individual misfit dislocations (MD) occur, on average, every 1.6 ± 0.35 nm laterally. To estimate areal dislocation density we need to consider the symmetry of the two films: the substrate sapphire belongs to rhombohedral family and its c-plane has a hexagonal 6-fold symmetry. TiN has a rocksalt cubic structure and the (111) plane of TiN has a 3-fold symmetry. The out-of-plane epitaxial relation of (111) TiN parallel to (001) $Al_2O_3$ forms two possible twins of TiN: one with $[1\bar{1}0]$ of TiN parallel to $[1\bar{1}0]$ of $Al_2O_3$ in-plane, and the other with $[0\bar{1}1]$ of TiN parallel to $[1\bar{1}0]$ of $Al_2O_3$ in-plane. Therefore, we expect similar MD profiles along every A-axis of the sapphire, i.e., every 60° from the current viewing zone axis of Figure 1d,e, but these additional profiles are not resolvable from a single view due to the depth resolution. This symmetry would imply three families of MD networks with similar spacings, suggesting a dislocation density of $6.8 \times 10^{13}$ $cm^{-2}$ at the planar interface of TiN/sapphire. A high density of MD is expected due to the large ~7.6% lattice mismatch between TiN and sapphire along the viewed zone axis. Fig. 2f illustrates the Fourier-filtered image with two MDs at the interface that are separated by 11 TiN atomic columns (white).

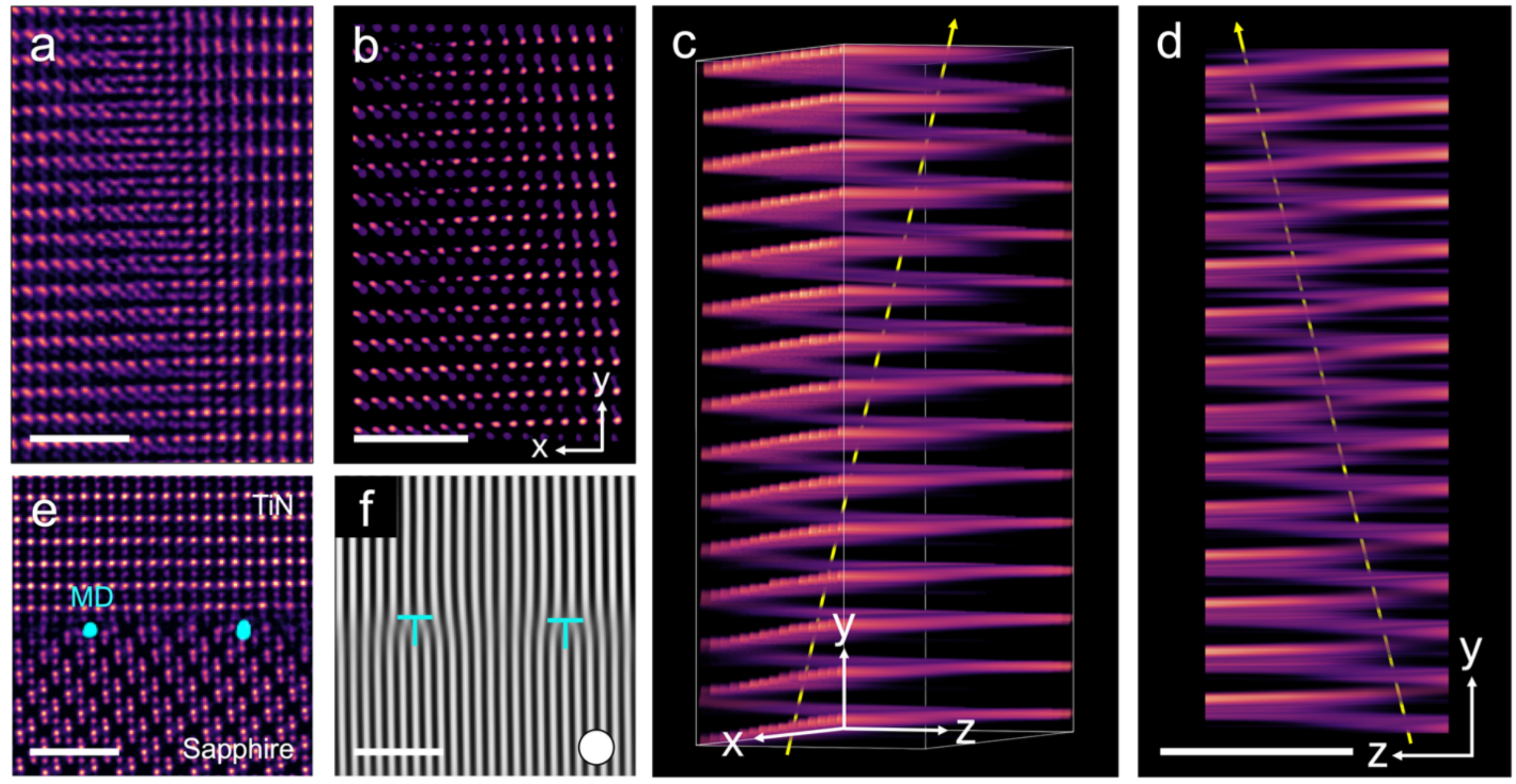

**Figure 2.** Individual screw- and edge-type dislocations in TiN/$Al_2O_3$. a) Screw dislocation in TiN from a single MEP slice reveals the characteristic zig-zag atomic arrangement. b-d) 3D visualization of the screw dislocation using multiple MEP slices: b) Projected view along the depth (z-)axis. Bright atomic rows are Ti atoms in the foreground slices. The darker atomic rows are also Ti atoms, but in the background slices that are deeper into the sample and have shifted as a result of the screw dislocation. c) Rotated view of (b) that reveals the 3D helical structure of a pure screw-type dislocation. Yellow arrow indicates the dislocation core. d) Alternate rotated view of (b), projected onto the x-axis. e) Edge dislocations due to lattice mismatch at the interface (i.e., misfit dislocations) give rise to a large geometric strain value in the periodic lattice image. The strain fields are determined via phase lock-in analysis[30] and indicated by cyan markers. The atomic image is a single 0.85 nm-thick slice of the full multislice reconstruction. f) Fourier-filtered image of (e) using $2\bar{2}0$ TiN and $1\bar{1}0$ sapphire Bragg peaks. The core of edge dislocations are indicated by cyan "T" markers. The white circle denotes the coarsening length associated with the size of the Gaussian mask in the Fourier-filtering process[30]. All scale bars are 1 nm.

We systematically identify the in-plane edge characteristics of dislocations using geometric phase lock-in[30] analysis. Similar to the well-known geometric phase analysis (GPA)[28,29], the phase lock-in algorithm calculates the geometric strain in an atomic-resolution image caused by local lattice distortions. At locations of defects such as edge dislocations, the discontinuity in the periodicity generates a singularity in local strain value, damped by the filter resolution to a large but finite value [Supporting Information Figure 3]. These large strain excursion points are then used to track edge-type defects in the atomic image as shown by cyan markers in Fig. 2e. This 2D analysis can be carried out over the entire 3D stack of the MEP reconstruction to create a 3D map of dislocations. Fig. 1c shows a schematic of the types of dislocations that can be identified using this technique: interfacial edge dislocations (i.e., MDs) are illustrated in cyan, SD in green, and the mixed-type threading dislocation (TD) in yellow. The experimental 3D mapping of these defects is displayed in Figs. 1f and 3a,b. The cyan MD markers, when interpolated along the depth (z-)direction becomes a series of strips that follows the trajectory of the MDs through the sample. Discontinuities of these trajectories often coincide with the conversion to a SD or TD, allowing the dislocation to propagate into the film bulk. The yellow markers map the trajectory of the edge-character of a TD, which forms where a MD disappears in Fig. 3b. Green markers indicate the strain singularities associated with lattice distortions from SDs.

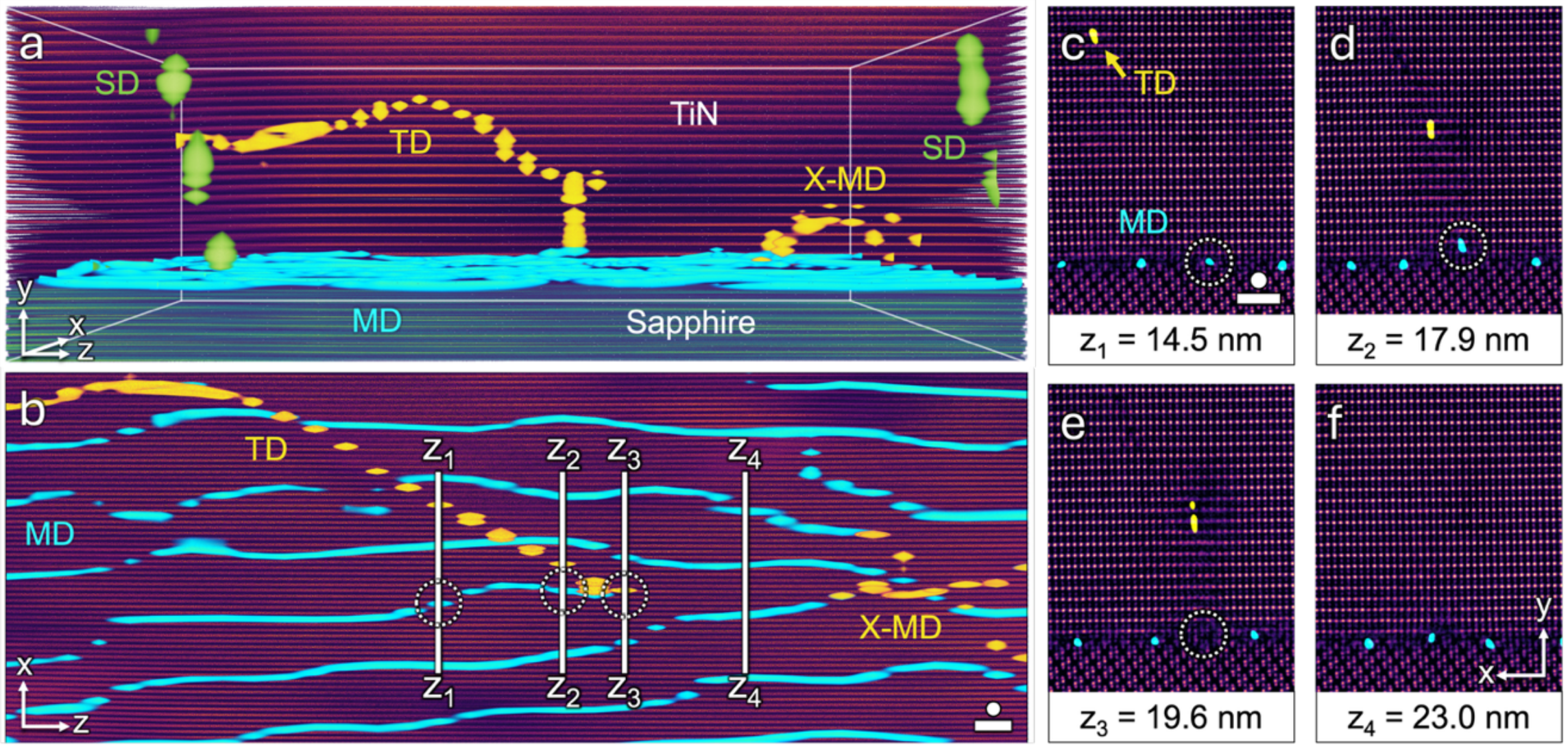

**Figure 3.** 3D mapping of individual, atomic-scale dislocations: misfit dislocation (MD), screw dislocation (SD), threading dislocation (TD), and a miscellaneous "crossing" misfit dislocation (X-MD). a) Side view and b) top view of the MEP reconstruction and mapped dislocations corresponding to Fig. 1f. Purple background is the atomic contrast from TiN and blue region below is the sapphire substrate. Cyan markers track the MDs present at the film-substrate interface due to lattice mismatch. Green markers indicate significant lattice distortions due to SDs. Yellow markers denote strong edge-characters present above the interface. Namely, a TD arises from a discontinued MD at the rightmost dotted circle in (b). The dotted circles in (b-e) track the same MD at depths $z_1$, $z_2$, and $z_3$ as it interacts with the TD. Further into the sample, an unexpected dislocation behavior is also observed: two MDs rise above the interface, crosses over, lowers, and continues along the depth as typical MDs. This type of "crossing" dislocation (labeled X-MD) is distinct from TD as the Burgers vectors remain unchanged during the crossover. c-f) MEP slices corresponding to linecuts at $z_1$-$z_4$ indicated in (b): c) At $z_1$, the MDs are more or less equally spaced according to the lattice mismatch while a TD hovers ~5 nm above. d) At $z_2$, a distinct TD in the TiN film approaches the substrate interface and possesses an edge-type Burgers vector antiparallel to the MDs; see [Supporting Information Figure 5]. e) After the annihilation of TD and the circled MD in (d), a residual lattice distortion is picked up as a strain signal above the empty dotted circle. f) The MD circled in (c,d) is no longer present and the neighboring MDs have shifted accordingly to maintain the lattice mismatch spacing. All scale bars are 1 nm and coarsening lengths are denoted by the white circles[28,29].

The side view (perpendicular to x-axis) and top-down view (projection along y-axis) in Figs. 3a and b highlight the behavior of the dislocations along the depth direction of the reconstruction. In particular, Fig. 3b and the corresponding plan-view slices in Fig. 3c-f at depths $z_1$ to $z_4$ show a single line of MD that disappears. In its place, an edge-type dislocation emerges and propagates toward the sample surface where it eventually terminates. This propagating TD is distinct from MD as it rises ~5 nm above the interface as shown in Fig. 3a, and possesses a Burgers vector antiparallel to the MDs; see [Supporting Information Figure 5]. The 3D mapping reveals another interesting dislocation behavior as well: two misfit dislocations approach one another (marked X-MD), rise and cross over, then continue as typical MDs in their switched positions. Unlike the TD, these "crossing" edge dislocations maintain identical Burgers vectors to that of the original MDs. The crossing is in the vicinity of a very-low-angle grain boundary, and may have formed as a stress-relief mechanism[53,54] along that boundary.

In general, geometric strain analysis must be approached and interpreted with caution. Historically applied to HRTEM images, GPA can evaluate the lattice strains of noisier images as long as the underlying periodicity is uniform. With these methods, dislocations can be imaged but little can be learned about the atomic structure, especially in complex systems with more than one atomic species due to limited spatial resolution and atomic number Z-contrast[28,29,55]. Applying GPA to STEM images is not uncommon and can provide insights about local lattice distortions with sub-nm resolution[56], but artifacts such as scan noise and drift pose challenges during the analysis or may output noisy strain signals. For both HRTEM and STEM, if the structure changes in depth, the projected image is likely unsuitable for quantitative GPA. With a well-converged, position-corrected MEP reconstruction the analysis should be more robust. High SNR atomic resolution and depth-sectioned slices provide well-behaved strain signals in GPA and likewise in phase lock-in analysis. It is, however, worth emphasizing that GPA and phase lock-in analyses are purely geometry-based strain calculations and are agnostic to the defect type. The calculated strain field is also directly dependent on the orientation and SNR of the selected peaks in the FFT. For example, selecting a Bragg peak that displays vertical lattice planes in the inverse Fourier transform (e.g., Fig. 2f) is well-suited for capturing horizontal displacement fields, which are prominent in edge-type MDs with horizontal Burgers vectors (Supporting Information Figure 3c). Horizontal displacement fields, however, can also be present in SDs that possess primarily vertical Burgers vectors. The SDs marked in Figs. 1f and 3a are such "edge-type" signals originating from the projected 3D strain field of a SD, demonstrating that strain field in one specific orientation cannot unambiguously determine the defect type. In this particular case, selecting a Bragg peak along a different orientation may provide clearer signals when systematically identifying SDs as demonstrated in [Supporting Information Figure 4]. In general, dislocations introduce 3D strain fields and a 2D strain analysis alone is not sufficient to assign defect types. By extracting strain fields along multiple orientations and analyzing through a stack of depth-resolved 2D images, we can more accurately assign the dislocation type based on their strain orientation and 3D evolution. With sub-Ångstrom resolution reconstruction from MEP, we can further verify the analysis by directly examining the 3D atomic structure.

Even with MEP, further care must be exercised during the phase analysis. First, a noisy atomic image, as in the case of low-dose MEP reconstructions, will result in a noisy (virtual) diffraction pattern when Fourier-transformed. This is an issue as tight Fourier-filtering of noise may generate seemingly plausible yet false periodic structure. Secondly, GPA and phase lock-in require a well-defined underlying periodicity to "lock-in" to. A large-scale change in the structure, such as a crack[22] or a broken or amorphous sample area without a well-defined lattice will result in artificial strain signals at and around those regions. Finally, it is important to keep in mind the loss of real-space resolution depending on the size of the Fourier filter mask[30]. A larger mask in Fourier-space will correspond to a higher resolution of the Fourier-filtered features in real-space, but will also incorporate more noise. A smaller filter size can suppress noise, but the real-space strain resolution will be diminished which manifests in real-space as larger apparently-ordered domains. It is always important to be aware of this bandwidth-limiting real-space coarsening length[30]. This resolution limit is indicated by white circles in Figs. 2 and 3; any strain features varying rapidly at a scale smaller than this length cannot be interpreted reliably. [See also Supporting Information Figure 3].

In summary, we achieved atomic-scale 3D mapping of individual structural defects within a high-density dislocation network in an epitaxial TiN/sapphire heterostructure using multislice electron ptychography. With lateral information transfer limit of 0.33 Å and depth resolution of 3.7 nm, we systematically identified and distinguished multiple dislocation types: edge-type misfit, pure screw, and mixed-type threading dislocations. By visualizing their 3D arrangements, the transition from misfit to threading dislocation was observed at the atomic scale, as well as an unconventional behavior of two crossing misfit dislocations. Importantly, this capability was achieved from a single scan acquired in ~15 seconds, covering a 15 nm × 15 nm field-of-view (although only 15 nm × 9 nm was reconstructed in this work for computational efficiency) and ~40 nm depth. For MEP conditions optimized for semiconductor devices, field-of-views ranging from 20 nm × 20 nm (with ~40 nm thickness) to 40 nm × 40 nm (with ~20 nm thickness) can be acquired in just 7 seconds[37]. Efficient acquisition allows for multiple MEP datasets taken in rapid succession and enable several reconstructions to be carried out in parallel – a capability that is limited only by availability of computational resources. Continued advances in detector technology[41] will further accelerate these acquisition times, and highly-efficient algorithm implementations[57,58] have already reduced the reconstruction time from several days to a few hours, with additional improvements to be expected from Moore's law alone. In addition to increasing throughput, these developments make it practical to reconstruct substantially larger fields-of-view in samples up to ~50 nm thick[37,48], overcoming computational limitations that previously constrained MEP to relatively small sample areas and thin samples. As material growth and fabrication processes continue to improve, device sizes will decrease, and so may defect densities. This will require larger imaging areas at high resolution for statistically meaningful characterization, which should be achievable with the expected advances in automated acquisitions and processing.

## ASSOCIATED CONTENT

**Supporting Information**

The following files will be available free of charge upon publication:

- Supporting Information: Experimental details, methods, and supplementary data (.pdf)
- EMPAD-G2 ptychography data (.h5)
- PtyRAD reconstruction parameter file (.yml)
- Reconstructed image stack (.tif)
- Reconstructed model (.pt)

AUTHOR INFORMATION

**Corresponding Author**

*david.a.muller@cornell.edu

**Author Contributions**

E.C.C. acquired and analyzed the STEM and MEP data and performed the MEP reconstruction via PtyRAD. C.-H.L. and S.Z. aided the reconstruction, supervised by D.A.M. A.I. grew the TiN films. K.-Y.P. and C.S.C. prepared the STEM sample lamella at the request of the collaboration. N.P. provided early guidance and training to E.C.C. and engaged in meaningful discussions about the result. D.J. and D.A.M. led the collaborative project funded by AFOSR and LPS. This manuscript was written by E.C.C. and reviewed by all authors.

**Funding Sources**

This work was supported by the Air Force Office of Scientific Research under award number FA9550-23-1-0688. Sample fabrication was carried out at the Cornell NanoScale Facility, an NNCI member supported by NSF grant NNCI-2025233. This work also used facilities of the Platform for the Accelerated Realization, Analysis, and Discovery of Interface Materials (PARADIM) and Cornell Center for Materials Research (CCMR) supported by NSF DMR-2039380 and DMR-1719875. FIB sample preparation was also supported by GRDC (Global Research Development Center) Cooperative Hub Program through the National Research Foundation of Korea (NRF) funded by the Ministry of Science and ICT (MSIT) (RS-2023-00258359).

## Acknowledgements

E.C.C. would like to thank N. Schnitzer for invaluable assistance and feedback in using the KEMSTEM package (https://github.com/noahschnitzer/kemstem) for phase lock-in analysis. E.C.C. would also like to thank S. Karapetyan for providing guidance in using the Tomviz software (https://tomviz.org/) to create the 3D visualizations. The resonator device was fabricated by H. Lu under the supervision of V. Fatemi. We thank J. Grazul, P. Carubia, M. Thomas and M. Silvestry Ramos for training and support with the microscopes.

## Data Availability

Data and reconstruction parameters will be available upon publication.

## Abbreviations

3D three-dimensional; 2D two-dimensional; (S)TEM (scanning) transmission electron microscopy; DF dark-field; HRTEM high-resolution transmission electron microscopy; ADF annular dark-field; MEP multislice electron ptychography; SNR signal-to-noise ratio; MD misfit dislocation; SD screw dislocation; TD threading dislocation; X-MD "crossing" misfit dislocation; GPA geometric phase analysis.